\documentclass[prl,twocolumn,superscriptaddress,showpacs,a4paper]{revtex4}

\usepackage{times} \normalfont \usepackage[T1]{fontenc}
\usepackage{graphicx}

\def\ealla#1{{\rm e}^{#1}}

\begin{document}

\title{Structure, rotational dynamics, and superfluidity 
of small OCS-doped He clusters}

\author{Saverio Moroni} \email{moroni@caspur.it} \affiliation{{\sl
  SMC} INFM --
  Istituto Nazionale per la Fisica della Materia and Dipartimento
  di Fisica, Universit\`a di Roma {\sl La Sapienza} \\ Piazzale Aldo
  Moro 2, I-00185 Rome, Italy}

\author{Antonio Sarsa} \email{sarsa@sissa.it} \altaffiliation{Present
  address: Departamento de Fisica Moderna, Universidad de Granada,
  E-18071 Granada, Spain.} \affiliation{SISSA and INFM {\sl
    DEMOCRITOS} National Simulation Center \\ Via Beirut 2-4, I-34014
  Trieste, Italy}

\author{Stefano Fantoni} \email{fantoni@sissa.it} \affiliation{SISSA
  and INFM {\sl DEMOCRITOS} National  
  Simulation Center \\ Via Beirut 2-4, I-34014 Trieste, Italy}

\author{Kevin E. Schmidt} \email{Kevin.Schmidt@asu.edu}
\altaffiliation{Permanent address: Department of Physics and
  Astronomy, Arizona State University, Tempe, Arizona 85287.}
\affiliation{SISSA and INFM {\sl DEMOCRITOS} National Simulation
  Center 
  \\ Via Beirut 2-4, I-34014 Trieste, Italy}

\author{Stefano Baroni} \email{baroni@sissa.it}
\altaffiliation{Permanent address: SISSA and DEMOCRITOS, Trieste,
  Italy.} \affiliation{SISSA and INFM {\sl DEMOCRITOS} National 
  Simulation Center \\ Via Beirut 2-4, I-34014 Trieste, Italy} 
\affiliation{Chemistry Department, Princeton University, Princeton NJ
  08544, USA}

\date{December 11, 2002}

\begin{abstract}
  The structural and dynamical properties of OCS molecules solvated in
  Helium clusters are studied using reptation quantum Monte Carlo, for
  cluster sizes $n=3-20$ He atoms. Computer simulations allow us to
  establish a relation between the rotational spectrum of the solvated
  molecule and the structure of the He solvent, and of both with the
  onset of superfluidity. Our results agree with a recent
  spectroscopic study of this system, and provide a more complex and 
  detailed microscopic picture of this system than inferred from
  experiments.
\end{abstract}


 
\maketitle

Solvation of atoms and molecules in He nanodroplets provides a way to
study their properties in an ultra-cold matrix, and also offers a
unique opportunity to probe the physics of quantum fluids in confined
geometries. Research in this field has been recently reviewed in Ref.
\cite{clss01}, with emphasis on experiment and hydrodynamic modeling,
and in Ref. \cite{khpbw00}, with emphasis on computer simulations.

Carbonyl sulfide (OCS) is one of the most widely studied dopants of He
clusters (OCS@He$_n$) \cite{jaeger,pgw01a}, both because of its strong
optical activity in the infrared and microwave spectral regions, and
also because the OCS@He$_1$ complex is spectroscopically well
characterized \cite{Higgins,Tang}, thus providing a solid benchmark
for the atom-molecule interaction potential which is the key
ingredient of any further theoretical investigation.

In the quest of fingerprints of superfluidity in the spectra of small-
and intermediate-size He clusters, Tang et {\sl al.} have recently
determined the vibrational and rotational spectra of OCS@He$_n$ at
high resolution for $n=2- 8$ \cite{jaeger}. The main results of that
investigation are: {\sl i)} the rotational constant of the solvated
molecule roughly equals the nanodroplet (large-$n$) limit \cite{gtv98}
at $n=5$, but then undershoots this asymptotic value up to the maximum
cluster size ($n=8$) attained in that work; {\sl ii)} the centrifugal
distortion constant has a minimum at $n=5$, thus indicating that
the complex is more rigid at this size; {\sl iii)} the fundamental
vibrational frequency of OCS is not a monotonic function of the number
of He atoms, but it displays a maximum at $n=5$, again suggesting a
stronger rigidity at this size. Findings {\sl ii)} and {\sl iii)}
suggest---and our study of the rotational spectra confirms---that
$n=5$ is a magic size related to the structure of solvent atoms around
the solvated molecule. Finding {\sl i)} implies the existence of a
(yet to be determined) minimum in the rotational constant as a
function of the cluster size. The occurrence of this minimum was
interpreted as due to quantum exchanges which would decrease the
effective inertia of the first solvation shell and would thus be a
signature of the onset of {\sl superfluidity}\ in this finite system
\cite{jaeger}.

Quantum simulations are complementary to the experiment for
understanding the properties of matter at the atomic scale. In fact,
while being limited by our incomplete knowledge of the inter-atomic
interactions and by the size of the systems one can afford to examine,
simulations provide a wealth of detailed information which cannot be
obtained in the laboratory. In this paper we show how state-of-the art
quantum Monte Carlo simulations can be used to get an unparalleled
insight into the exotic properties of small He clusters. At the same
time, some of the dynamical properties thus predicted are so sensitive
to the details of the atom-molecule interactions, to provide a very
accurate test of the available model potentials.

The structure and the rotational dynamics of OCS@He$_n$ clusters are
studied here in the small-to-intermediate size regime ($n=3-20$),
using {\sl reptation quantum Monte Carlo} (RQMC) \cite{noantri}, a
technique by which ground-state properties (such as the density
distribution and various static and dynamic correlations) can be
determined with high accuracy. In particular, the analysis of the
dynamical dipole correlation function allows us to resolve, for the
first time theoretically, several rotational components of the
spectrum of an interacting quantum system. RQMC combines the best
features of the path-integral (PIMC) and diffusion quantum Monte Carlo
(DQMC) techniques, yet avoiding some of their drawbacks: RQMC is not
affected by the systematic errors which plague the estimate of
ground-state expectation values in DQMC (the so called mixed-estimate
and population-control biases) and gives easy access to time
correlations, from which dynamical properties can be extracted
\cite{noantri}; at the same time, the RQMC technique is specially
tailored to the zero-temperature (ground-state) regime, where PIMC
becomes inefficient.

Both the He-He and the He-OCS interactions used here are derived from
accurate quantum-chemical calculations \cite{korona,hoho01}. The OCS
molecule is allowed to perform translational and rotational motions,
but it is assumed to be rigid. The trial wavefunction is chosen to be
of the Jastrow form: $ \Phi_T = \exp \left [ -\sum\limits_{i=1}^n
{\cal U}_1(r_i,\theta_i) -\sum\limits_{i<j}^n {\cal U}_2(r_{ij})
\right ], $ where $r_{ij}$ is the distance between the i-th and the
j-th helium atoms, $r_i=|{\bf r}_i|$ the distance of the i-th atom
from the center of mass of the molecule, and $\theta_i$ is the angle
between the molecular axis and ${\bf r}_i$. On account of the high
anisotropy of the He-OCS potential, ${\cal U}_1$ is expanded as a sum
of five products of radial functions times Legendre polynomials. All
radial functions (including ${\cal U}_2$) are optimized independently
for each cluster size with respect to a total of 27 variational
parameters. The propagation time  is set to $\rm 1~
^oK^{-1}$, with a time step of $\rm 10^{-3}~ ^oK^{-1}$. The
effects of the length of the time step and of the projection time 
\cite{noantri} have
been estimated by test simulations performed by halving the former or
doubling the latter. These effects were barely detectable on the total
energy, and negligible with respect to the statistical noise for the
properties on which we base our discussion below.

\begin{figure}
  \hbox to \hsize{\hfill
    \includegraphics[height=95mm]{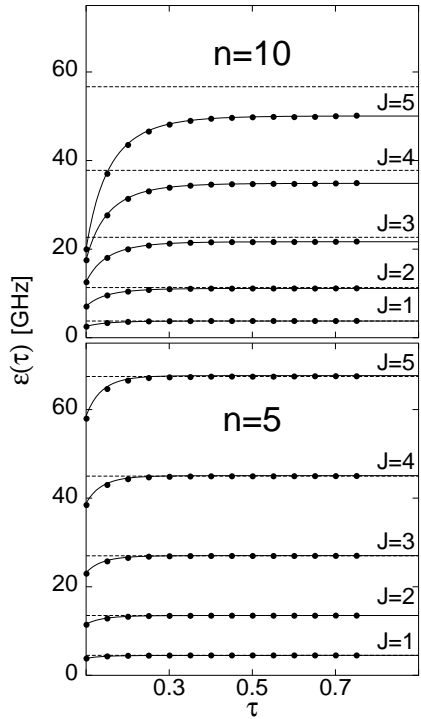}
    \hfill
  }
  \caption{
    Logarithm of the multipole correlation function, normlized by the
    spectral weight of the lowest-lying state: $\epsilon(\tau)=-\log{ 
    \left (c_J(\tau)/A_J \right ) / \tau }$, for $n=5$ 
    (lower panel) and $n=10$ (upper panel). Time units are $\rm
    ^oK^{-1}$.
  }
\vspace{-5mm} 
\end{figure}

The absorption spectrum of a molecule solvated in a non polar
environment is given by the Fourier transform of the autocorrelation
function of its dipole, $\bf d$: $\quad I(\omega) \propto 2\pi \sum_n
| \langle \Psi_0|{\bf d}|\Psi_n \rangle |^2 \delta(E_n-E_0-\omega) =
\int \ealla{i\omega t} \langle {\bf d}(t) \cdot {\bf d} (0) \rangle
dt$, where $\Psi_0$ and $\Psi_n$ are ground- and excited-state
wavefunctions of the system respectively, $E_0$ and $E_n$ the
corresponding energies, and $\langle . \rangle$ indicates ground-state
expectation values. The dipole of a linear molecule---such as OCS---is
oriented along its axis, so that the optical activity is essentially
determined by the autocorrelation function of the molecular
orientation versor: $c(t)= \langle {\bf n}(t)\cdot {\bf n}(0)
\rangle$. Within RQMC the analytical continuation to imaginary time of
this correlation function, $\bar c(\tau) = c(-i\tau)$, can be
conveniently calculated without any other approximations than those
implicit in the use of a given parametrization of the interatomic
potentials \cite{noantri}. From now on, when referring to {\sl time
correlation functions}, we will mean quantum correlations in imaginary
time, and will we drop the bar over $\bar c(\tau)$. Analytical
continuation to imaginary time transforms the oscillatory behavior of
the real-time correlation function---which is responsible for the
$\delta$-like peaks in its Fourier transform---into a sum of decaying
exponentials whose decay constants are the excitation energies,
$E_n-E_0$, and whose spectral weights are proportional to the
absorption oscillator strengths, $|\langle \Psi_0|{\bf d}|\Psi_n
\rangle |^2$. Dipole selection rules imply that only states with $J=1$
can be optically excited from the ground state. Information on excited
states with different angular momenta, $J$, can be easily extracted
from the multipole correlation functions, $c_J(\tau)$, defined as the
time correlations of the Legendre polynomials: $c_J(\tau) = \left
\langle P_J ( {\bf n}(\tau) \cdot {\bf n}(0) ) \right \rangle \equiv
\left \langle {1\over 2J+1} \sum_{M=-J}^J Y^*_{JM}\bigl ({\bf n}(\tau)
\bigr ) Y_{JM}\bigl ({\bf n}(0) ) \right \rangle$. For each value of
$J$, the value of the lowest-lying excitation energy, $\epsilon_J$
---{\sl i.e.} the smallest decay constant in $c_J(\tau)$---as well as
the corresponding spectral weight, $A_J$, can be extracted from a fit
to $c_J(\tau)$. We have verified that the values so obtained are rather
insensitive to the details of the fitting procedure.

In Fig. 1 we display multipole correlations calculated up to $J=5$ for
OCS@He$_{5}$ and OCS@He$_{10}$. For a rigid-top, time correlations are
described by a single exponential for each value of $J$, $c_J(\tau) =
\ealla{-\epsilon_J\tau}$, whose decay constant is $\epsilon_J=B
J(J+1)$. The rigid-top energy levels are displayed as dashed
lines. Time correlations deviate from a single-exponential behavior
only at short times, thus indicating that the non rigidity of the
cluster only affects the spectra at very high frequencies, the
deviation being larger the larger the angular momentum. The dependence
of $\epsilon$ on $J$ is accurately predicted by the rigid-top model
for $n=5$, while it is less so for $n=10$. In all cases the excitation
energies are well described by a formula containing an effective
centrifugal distortion constant: $\epsilon_J=B J(J+1) -DJ^2(J+1)^2 $
\cite{nota-correlata}.

\begin{figure}
  \hbox to \hsize{\hfill
    \includegraphics[height=56mm]{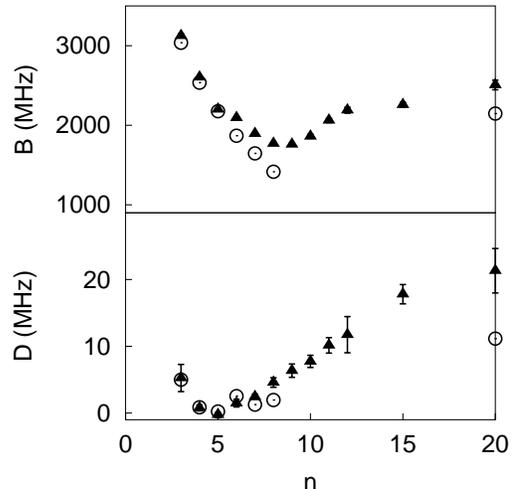}\vspace{5mm}
    \hfill
  }
  \caption{Rotational constants (upper panel) and rotational
    distortion constant (lower panel, see text) of OCS@He$_n$, as
    functions of the cluster size, $n$.  Triangles indicate the results
    of the present simulation, whereas circles are experimental data
    from Ref. \cite{jaeger}. 
  }
\end{figure}

In Fig. 2 we display the rotational constant, $B$, and centrifugal
distortion constant, $D$, as functions of the cluster size. $B$
displays a minimum at $n=8- 9$, while $D$ is minimum at $n=5$,
indicating a greatest rigidity of the He@OCS complex at this size. In
order to discuss these findings and their relation with the structure
of the cluster, with the onset of superfluidity, and with the quality
of the interaction potential used for the simulation, we first define
$\rho({\bf r})$ as the He number-density distribution, and $\phi({\bf
r})$ as the ground-state expectation value of molecule-atom angular
motion correlation: $\phi({\bf r}) = - \left \langle \Psi_0 | {\bf L}
\cdot {\bf l}({\bf r}) | \Psi_0\right \rangle$. In this expression
${\bf L}$ indicates the angular momentum of the OCS molecule and ${\bf
l}({\bf r})$ is the angular-current operator of He atoms at point $\bf
r$: $ {\bf l}({\bf r}) = -i~ \hbar~{{\bf r} } \times \sum_l \left (
{\partial \over \partial \bf r_l} \delta({\bf r} - {\bf r}_l)+
\delta({\bf r} - {\bf r}_l) {\partial \over \partial \bf r_l} \right )
$.  The angular correlation function, $\phi({\bf r})$, is non-negative
and it is large whenever the angular motion of He atoms at point $\bf
r$ is strongly correlated with the rotation of the solvated molecule,
thus contributing to the molecular effective moment of inertia.

In Fig. 3 we display $\rho({\bf r})$ and $\phi({\bf r})$, as
calculated for cluster sizes $n=5,8,10,15$. The He density reaches its
maximum where the He-OCS interaction potential is minimum, {\sl i.e.}
in a {\sl doughnut} surrounding the molecule perpendicular to the OC
bond. This doughnut can accommodate up to 5 He atoms which rotate
rather rigidly with the OCS molecule, as demonstrated by the large
value of the angular correlation function. Each He atom within the
doughnut gives a same contribution to the cluster moment of inertia,
resulting in a constant slope of $B$ vs.  $n$ for $n\le 5$.

Secondary minima of the He-OCS potential exist on the molecular axis
near the two poles, the one nearest to the S atom being deeper. For
$n>5$ He atoms spill out of the region of the main minimum and spread
towards other regions of space, preferentially near the molecular
poles. A detailed analysis of the He density plots for $n=6,7,8$ shows
that---due to quantum tunneling to and from the doughnut---both polar
regions start to be populated as soon as the doughnut occupation is
completed ({\sl i.e.} for $n>5$). The number of Helium atoms near the
sulfur pole is 0.2, 0.3, and 0.7, for $n=6,7,8$, and 0.5, 0.8, and 1.1
near the oxygen pole. Helium density is larger near the oxygen
pole---rather that near sulfur where the potential is more
attractive---because a smaller energy barrier and a smaller distance
from the absolute minimum in the doughnut make quantum tunneling
between the absolute and the secondary minima easier in this case.

Quantum tunneling is the key for understanding the onset of
superfluidity, as well as the sensitivity of rotational spectra to the
details of the He-OCS interaction. For $n=6,7,8$ a sizeable He density
is found not only near the molecular poles, but also in the angular
region between the doughnut and the oxygen pole, where the energy
barrier is small. Non negligible angular correlations indicate that
atoms in this region (and not only near the poles) contribute to the
effective moment of inertia. For $n>8$ He atoms start to fill the
angular region between the doughnut and the S pole (see the $n=10$
panel of Fig. 3). The closure of He rings in the sagittal planes makes
atomic exchanges along these rings possible, thus triggering the same
mechanism which gives rise to superfluidity in infinite systems
\cite{feynman}. The onset of superfluidity is responsible not only for
negligible angular correlations in this region, but also for the
decrease of these correlations in the doughnut region where the
largest contributions to the cluster inertia come from. This is
demonstrated in the bottom-right panel of Fig. 3, where one sees that
the maximum of the angular correlation in the doughnut region starts
decreasing for $n>8$. 

It is also interesting to examine the dependence upon cluster size of
the spectral weights, $A_J$. It turns out that, while for $J=1$ the spectral
weight displays a maximum at $n=4-5$ and a minimum at $n=8$, the
maximum at $n=4-5$ is not followed by any minimum for $J=5$. We
interpret the maximum of $A_J$ at $n=4-5$ as another manifestation of
the greatest rigidity of the He@OCS cluster at this size, while the
behavior for $n>5$ is determined by the number of solvent states
available to couple with the lowest-lying molecular rotation. This
number increases with the cluster size for $J=5$, while it has a
maximum at $n=8-9$ for $J=1$. These findings are consistent with the
emergence of superfluid behavior at cluster size, as characterized by
the reduction of low-lying excited states in the He matrix
\cite{feynman}.

This quantum tunneling makes the rotational constants for $n>5$ rather
sensitive to the height and width of the barriers between different
minima of the OCS-He interaction potential. On the other hand, the
details of the potential off the main minimum in the doughnut hardly
affect the spectra for $n\le 5$ and, in particular, that of OCS@He$_1$
which is used as a benchmark for the quality of the potential
\cite{hoho01}. Our simulations predict rotational constants which are
in excellent agreement with experiment for $n\le 5$, while for $n\ge
6$ the predicted values of $B$ are too large and display too small a
slope as a function of $n$. We believe that both these facts are due
to the smallness of the barriers between the main and the secondary
energy minima, which determines an excessive spill-out of He atoms off
the main minimum for $n=6$ and an excessive propensity towards atomic
exchanges for $n\ge 6$. We expect that larger energy barriers would
result in a better agreement between theory and experiment for $6\le
n\le 8$ and, possibly, in a shift of the predicted minimum of the
rotational constant towards larger sizes. As a qualitative test, we
have redone some of our simulations with a uniformly scaled potential
($V_{\rm He-OCS} \rightarrow \alpha V_{\rm He-OCS}$). Although not
very realistic, this transformation captures the essential
modifications to be done to the potential, {\sl i.e.} it modifies the
height of the barriers while leaving the location of the minima
unchanged. For $\alpha=1.2$, we find that the values of $B$ for $n\le
5$ are left practically unchanged, while the agreement with experiment
is much improved for $n=6,7,8$. The resulting minimum value of $B$
stays located near $n=8-9$. This lets us believe that our prediction of
the position of the minimum rotational constant as a function of the
cluster size should be rather robust with respect to possible
improvements of the He-OCS effective potential.

\begin{widetext}
$$ \vbox {\hsize =178mm \small
    \noindent FIG. 3: Upper left panel: He-OCS interaction
    potential. The atoms constituting the molecule are displayed as
    circles whose radius is the corresponding Van der Waals radius
    (sulphur on the left).
    Panels with red background: upper mid panel, He number density;
    lower mid panel, angular motion correlation function (see text).
    Colors go from red to violet as in a rainbow, as the value of the
    function increases.  Lower right panel: maximum of the angular
    motion correlation function, as a function of the cluster size.
  $$ \includegraphics[height=63mm]{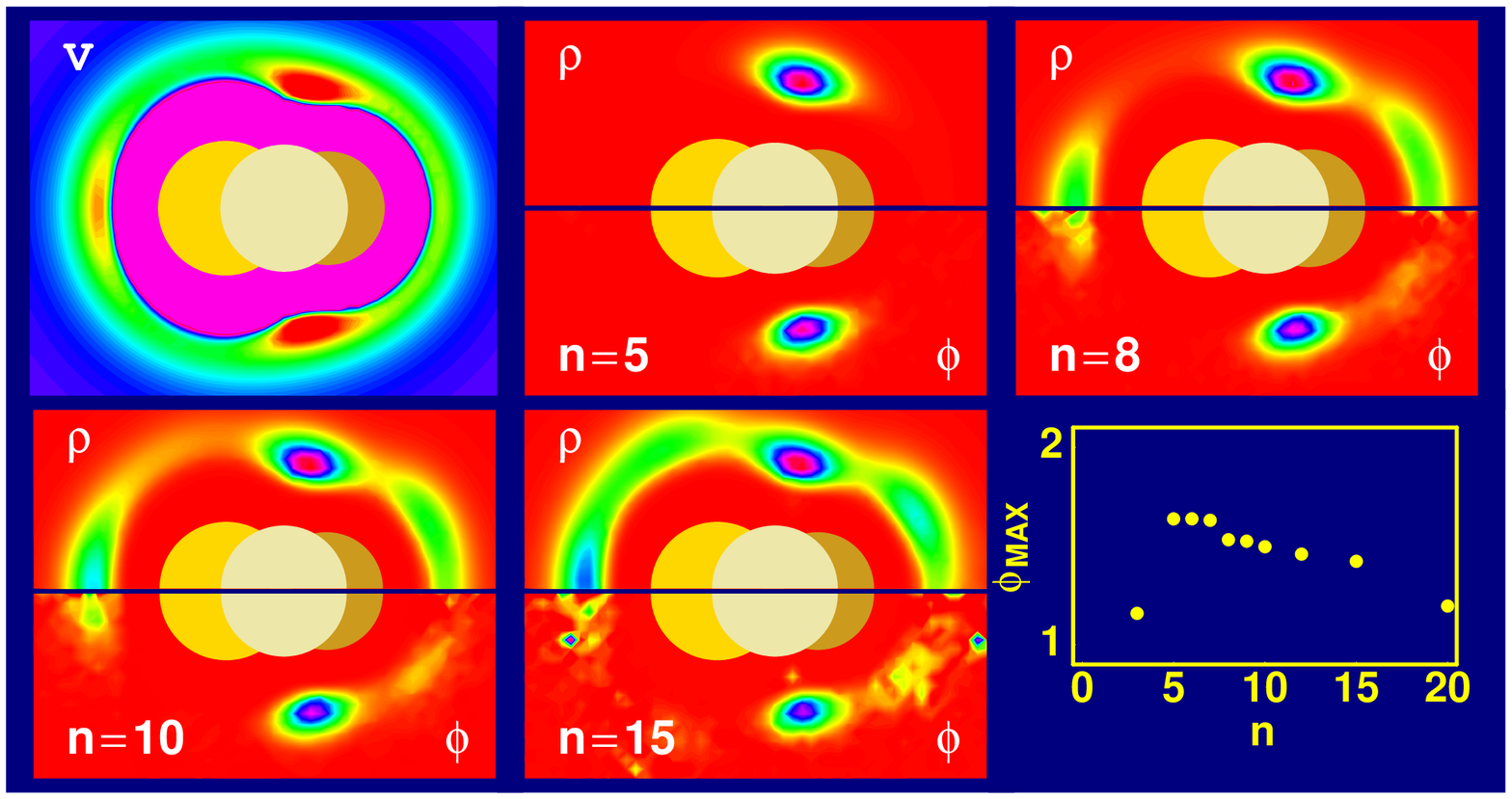} $$
} $$
\end{widetext}


In conclusion, we have shown that state-of-the art quantum Monte Carlo
simulation provide both a qualitative explanation of the relation
existing among structure, dynamics, and superfluidity in small He
clusters, and a sensitive test of the quality of atom-molecule
potentials, in regions which hardly affect the spectra of the
monoatomic molecular complex.
 
We are grateful to W. J\"ager for sending us a preprint of Ref.
\cite{jaeger} prior to publication, and to G. Scoles for bringing that
paper to our attention and for a careful reading of the first draft of
the present paper. Last but not least, we would like to thank both K.
Lehman and G. Scoles for many illuminating discussions held at the
Chemistry Department of the Princeton University where SB was the
grateful guest of R. Car while part of this work was being done.


\begin{thebibliography}{99}
\bibitem{nota-correlata} The evaluation of the distortion constant,
$D$, is much facilitated by the fact that large correlations exist
among the time series from which values of $c_J(\tau)$ are extracted
for different values of $J$ (they all derive from a same underlying
random walk). Because of this, the statistical error in the ratio of
any two $c_J$'s (whose logarithmic derivatives gives excitation
energies) is smaller than the error on individual $c_J$'s.
\end{thebibliography}
\end{document}